\documentclass[%
aps,
preprint,
amsmath,amssymb,
floatfix,
]{revtex4-2}

\usepackage{graphicx}
\usepackage{dcolumn}
\usepackage{bm}
\usepackage{etoolbox}
\usepackage{hyperref}

\graphicspath{{figures/}}

\begin{document}

\title{How Quasicrystals Remember: Hierarchical Memory Under Cyclic Shear}

\author{Edwin A. Bedolla-Montiel}
 \email{e.a.bedollamontiel@uu.nl}
\affiliation{Soft Condensed Matter \& Biophysics, Debye Institute for Nanomaterials Science, Utrecht University, Princetonplein 1, 3584 CC Utrecht, Netherlands.}

\author{Marjolein Dijkstra}
 \email{m.dijkstra@uu.nl}
\affiliation{Soft Condensed Matter \& Biophysics, Debye Institute for Nanomaterials Science, Utrecht University, Princetonplein 1, 3584 CC Utrecht, Netherlands.}

\date{\today}

\begin{abstract}
Quasicrystals occupy a unique middle ground between periodically ordered crystals and disordered
glasses, making them an ideal platform for examining the interplay between disorder and the emergence of mechanical memory. Using athermal quasistatic shear simulations, we show that two-dimensional dodecagonal quasicrystals  encode and recover memory under cyclic driving. Above the yielding transition, the response  becomes irreversible, characterized by  persistent shear bands and   locally transformed regions. 
Below yielding, cyclic shear with varying amplitudes produces a hierarchy of nested hysteresis loops in the stress-strain response characteristic of loop-return point memory.
By resolving the underlying reversible plastic events, we reveal  localized phason-like tile rearrangements as the elementary switching units and identify tile-switch hysterons  responsible for memory in the quasicrystal. Such a microscopic identification of the fundamental switching units is considerably more challenging, and often impossible, in amorphous solids. Despite their structural diversity, these rearrangements share  a compact core, sharp bistability, and an Eshelby-compatible elastic far field. In contrast, a periodic approximant of the quasicrystal lacks both the structural disorder and the bistable tile-switch rearrangements required for cyclic-shear memory, linking phason degrees of freedom to bistable hysterons.
\end{abstract}

\maketitle

Periodically driven materials can store information about their driving history~\cite{fiocco2013oscillatory,paulsenMechanicalMemoriesSolids2025,mungan2025self}. A prominent example is return point memory (RPM), first identified in ferromagnetic systems by Preisach~\cite{preisach1935magnetische}. RPM describes the ability of a system under cyclic driving to return exactly to previously visited states, thereby encoding information about its past.
The current theoretical framework attributes RPM to the collective behavior of bistable switching elements called hysterons~\cite{keimMemoryFormationMatter2019,munganStructureStateTransition2019a}. Under cyclic driving at different amplitudes, these hysterons switch hierarchically, generating nested hysteresis loops that enable the storage and retrieval of multiple memories.

So far, RPM has been studied primarily in disordered systems, such as amorphous solids~\cite{fiocco2013oscillatory,leishangthem2017yielding,berthier2025yielding} and crumpled sheets~\cite{bense2021complex,shohatMemoryCoupledInstabilities2022}. In amorphous solids under cyclic shear, for example, spatially correlated groups of particles undergo reversible rearrangements during loading and unloading, providing a microscopic mechanism for memory formation~\cite{fiocco2013oscillatory,fioccoEncodingMemorySheared2014,munganNetworksHierarchiesHow2019a,keimGlobalMemoryLocal2020}.
Periodic crystals are typically too ordered to support the localized, reversible rearrangements required for memory, whereas disordered glasses naturally exhibit rugged energy landscapes that promote bistable switching.
Quasicrystals occupy a unique intermediate position~\cite{zhaoAtomisticMechanismsDynamics2025}: they exhibit long-range orientational order while lacking translational periodicity~\cite{dotera2014mosaic}, and their mechanical response under cyclic shear has only recently begun to be explored~\cite{maireSelfassemblyQuasicrystalsCyclic2026}.
Indeed, their local structural environments are highly non-trivial and, in the limit of increasing rotational symmetry, approach a universal distribution whose spread of near-neighbor distances more closely resembles that of an amorphous structure rather than a periodic crystal~\cite{mendozasosaStructuralStudiesLocal2023}.
However, whether quasicrystals can support mechanical memory remains an open question.

Two-dimensional dodecagonal quasicrystals (DDQCs), formed by square-triangle tilings~\cite{oxborrowRandomSquaretriangleTilings1993,imperor-clercSquaretriangleTilingsInfinite2021}, can be stabilized by core-softened interaction potentials~\cite{kryuchkov2018complex,padillaPhaseBehaviorTwodimensional2020},
providing a well-controlled model for studying mechanical properties of quasicrystals.
These tilings host different types of defects~\cite{fayenQuasicrystalBinaryHard2024,ulugolDefectsEnhanceStability2024,ulugol2026vacancydefectssquaretriangletilings}, and bonds near such defects can switch between distinct local configurations~\cite{ulugol2026vacancydefectssquaretriangletilings}.
This suggests that DDQCs may support the localized, reversible rearrangements required for mechanical memory, while also providing a platform to investigate how preparation history influences the collective organization of these rearrangements.

In this work, we study the zero-temperature mechanical behavior of DDQCs under quasistatic conditions. We show that DDQCs undergo a yielding transition as the strain amplitude increases. At large strain amplitudes, shear bands emerge accompanied by local regions of transformed symmetry. In contrast, at small strain amplitudes, cyclic driving with varying amplitudes produces a series of nested hysteresis loops, characteristic of RPM.
Comparing samples prepared at lower and higher parent temperatures, we find that both sample sets satisfy the same macroscopic loop-return point memory (loop-RPM) criterion, while higher-temperature samples explore systematically broader reversible excursions.
By examining a representative event at finer strain resolution, we identify a localized phason-like tile rearrangement as the elementary switching unit, defining a tile-switch hysteron embedded within the quasicrystal responsible for mechanical memory.

\section*{Results}
We generate energy-minimized configurations of a 2D DDQC (\emph{SI Appendix}, Fig. S1)  using a core-softened interaction potential at a number density $\rho\sigma^2=0.94$ and  two preparation temperatures $k_BT/\epsilon=0.05$ and 0.15. The resulting configurations are subsequently subjected to an athermal quasistatic shear protocol (\emph{SI Appendix}, Fig. S2), see the \emph{ Materials and Methods} for further details. 

\subsection*{Yielding Transition}
\begin{figure}
    \centering
    \includegraphics[width=0.9\columnwidth]{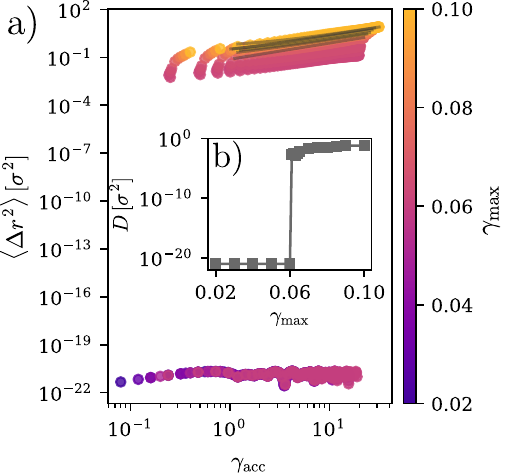}
    \caption{a) Mean squared displacement  MSD($\gamma=0$) of a 2D dodecagonal quasicrystal with $N=4096$ particles as a function of accumulated strain $\gamma_{\mathrm{acc}}$ for varying strain amplitudes $\gamma_{\mathrm{max}}$ as indicated by the color bar, after reaching steady state at $N_c=50$ cycles. Gray solid lines indicate exemplary linear fits used to extract the diffusion coefficient $D$ for all $\gamma_{\mathrm{max}}$, including sub-yield amplitudes where the fitted slope is indistinguishable from zero. At low $\gamma_{\mathrm{max}}$, the MSD remains close to zero, consistent with an elastic limit cycle, whereas at large $\gamma_{\mathrm{max}}$ it
    increases linearly, indicating plastic rearrangements. b) Diffusion coefficient $D$, extracted from linear fits to the MSD curves, showing an apparent yielding transition at $\gamma_y \approx 0.065$.}
    \label{fig:msd_diffusion}
\end{figure}

To locate the yielding transition, we sweep the strain amplitude over $\gamma_{\mathrm{max}}\in[0.02,0.1]$ in increments of $0.005$. For each amplitude, we perform up to $50$ cycles to reach a steady-state limit cycle, monitored via the energy per particle $U/N$ (\emph{SI Appendix}, Fig. S3).
We characterize the mechanical response through the virial shear stress $\sigma_{xy}$ (in units of $\epsilon/\sigma^2$) in  Fig. S3 in the \emph{SI Appendix}, and the stroboscopic mean squared displacement, defined as MSD$(\gamma)= \left\langle \sum_{i=1}^N |{\bf r}_i(\gamma_{\mathrm{acc}}+\gamma)-{\bf r}_i(\gamma_{\mathrm{acc}})|^2 / N \right\rangle$.
The MSD is evaluated at stroboscopic ($\gamma=0$) snapshots, averaged over independent samples, and plotted in Fig.~\ref{fig:msd_diffusion} as a function of the accumulated strain $\gamma_{\mathrm{acc}}=4N_c\gamma_{\mathrm{max}}$ at $\gamma=0$ with $N_c$ the cycle number.
At low $\gamma_{\mathrm{max}}$, the stress response is linear and the MSD remains close to zero, consistent with elastic behavior.
In contrast, at large $\gamma_{\mathrm{max}}$, hysteresis loops emerge in the stress-strain curves, and the MSD increases linearly with accumulated strain $\gamma_{\mathrm{acc}}$, indicating irreversible plastic rearrangements (see \emph{SI Appendix}, Fig. S3).
Both observations are consistent with previous results for amorphous solids~\cite{bhaumik2022yielding}.
Linear fits to the MSD (Fig.~\ref{fig:msd_diffusion}a) yield a diffusion coefficient $D$ that increases sharply near $\gamma_y \approx 0.065$ for $N=4096$ (Fig.~\ref{fig:msd_diffusion}b), thereby identifying the yielding transition.
Finite-size effects slightly shift this value (\emph{SI Appendix}).
At high strain amplitudes above the yielding transition, plastic rearrangements become irreversible and the DDQC develops persistent shear bands, accompanied by the emergence of locally transformed regions whose symmetry differs from that of the surrounding quasicrystal, as revealed by maps of the local bond order parameter in the \emph{SI Appendix}, Fig. S5.
This loss of quasicrystalline order provides a microscopic mechanism for the hysteresis loops observed for $\gamma\geq\gamma_y$, and is consistent with recent studies on cyclically sheared quasicrystals~\cite{parisi2017shear,bhaumik2022yielding,maireSelfassemblyQuasicrystalsCyclic2026}.

\subsection*{Loop-Return Point Memory}
Given that the DDQC exhibits plasticity and shear-band formation above $\gamma_y$, we next investigate whether mechanical memory can be encoded in samples subjected to cyclic shear below the yielding transition, where plastic rearrangements remain fully reversible.

\begin{figure*}
    \centering
    \includegraphics[width=\textwidth]{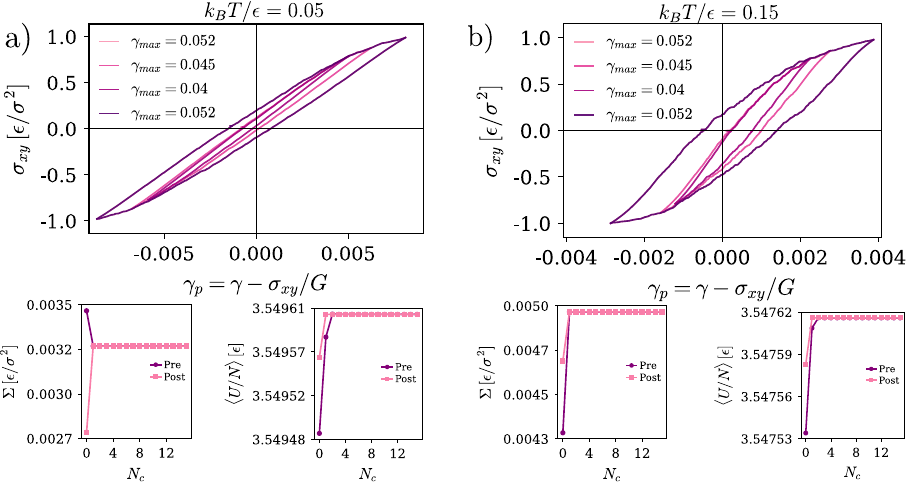}
    \caption{Hierarchical cyclic-shear response of a 2D dodecagonal quasicrystal with  $N=4096$ particles prepared at parent temperatures a) $k_B T/\epsilon=0.05$ and b) $k_B T/\epsilon=0.15$. The nested protocol follows $\gamma_3\to\gamma_2\to\gamma_1\to\gamma_2\to\gamma_3$, with $\gamma_3=0.052$, $\gamma_2=0.045$, $\gamma_1=0.04$, and $10$ cycles at each amplitude. For each temperature, the upper panel shows $\sigma_{xy}$ as a function of plastic strain $\gamma_p=\gamma-\sigma_{xy}/G$, where $G$ is obtained from a fit to the initial loading branch. The lower panels show the hysteresis-loop area $\Sigma \equiv \oint \sigma_{xy} d \, \gamma$ and the energy per particle $U/ N$ as a function of cycle number $N_c$ during both pre- and post-excursion cycles at $\gamma_{\max}=0.052$, showing convergence  to a  limit cycle. %
    }
    \label{fig:rpm-loop-both-temps}
\end{figure*}

Fig.~\ref{fig:rpm-loop-both-temps} illustrates the hierarchical protocol for a representative sample prepared at both $k_BT/\epsilon=0.05$ and $k_BT/\epsilon=0.15$, driven through the sequence $\gamma_3\to\gamma_2\to\gamma_1\to\gamma_2\to\gamma_3$, with $\gamma_3=0.052$, $\gamma_2=0.045$, $\gamma_1=0.04$, and $10$ cycles at each amplitude.
The resulting stress $\sigma_{xy}$ - strain $\gamma$ curves are clearly nested. However, the dominant elastic contribution $\sigma_{xy}\approx G\gamma$ compresses the sub-loops along the strain axis.
Replotting the shear stress against the plastic strain $\gamma_p = \gamma - \sigma_{xy}/G$, where $G$ is obtained from a fit to the initial loading branch, isolates the dissipative response and makes the loop hierarchy explicit.
For both temperatures, the system reaches its limit cycle within the first few cycles, as shown by the convergence of the mean energy per particle $\langle U/ N\rangle$ and the loop area $\Sigma = \oint \sigma_{xy} d \, \gamma$ in the bottom rows of Fig.~\ref{fig:rpm-loop-both-temps}.
To test how loop closure depends on preparation temperature, we analyze two independent sample sets of $25$ replicas, self-assembled at $k_{B}T/\epsilon=0.05$ and $k_{B}T/\epsilon=0.15$ and driven through an identical hierarchical loading schedule. The replicas are self-assembled and sheared independently, so each run is a separate realization of the protocol. For each run, we compare the normalized stress--strain loop of the last cycle at $\gamma_3$ before and after the excursion using the root-mean-square distance
\begin{equation}
  d^{\mathrm{return}}
  = \left[
      \frac{1}{N_\mathrm{grid}}
      \sum_{k=1}^{N_\mathrm{grid}}
      \bigl(\sigma_{xy}^{\,\mathrm{pre}}(\gamma_k)
            - \sigma_{xy}^{\,\mathrm{post}}(\gamma_k)\bigr)^{2}
    \right]^{1/2},
    \label{eq:dreturn}
\end{equation}
on a common strain grid. To quantify the intrinsic cycle-to-cycle variability of a converged limit cycle, we first subject the system to 10 identical shear cycles at the same strain amplitude $\gamma_3$. After the first one or two cycles, the configuration converges to a limit cycle, and the distance between consecutive cycles reaches a steady value. We define the baseline distance, $d^{\mathrm{baseline}}$, as the average distance between consecutive cycles over the last three cycle pairs. This baseline characterizes the numerical noise floor of the converged limit cycle. Because the cycle-to-cycle distance fluctuates slightly even after convergence, we use the largest observed baseline value as a conservative estimate of the intrinsic variability. A hierarchical loading protocol is therefore classified as exhibiting loop-RPM when the distance between the pre- and post-excursion configurations satisfies $d^{\mathrm{return}} \le \max(d^{\mathrm{baseline}}) + \epsilon_{\mathrm{mach}}$, where $\epsilon_{\mathrm{mach}}=10^{-12}$ is a machine-precision tolerance. This criterion ensures that the returned state is indistinguishable from the original limit cycle within its natural cycle-to-cycle fluctuations.

Every run satisfies the normalized stress-loop criterion, and the corresponding energy and overlap criteria are fulfilled across all replicas.
Pre- and post-excursion cycles differ only within the intrinsic cycle-to-cycle fluctuations of the converged block, while the overlap remains essentially perfect in both datasets.
Loop-RPM is therefore robust at both parent temperatures.
The hierarchical loop structure persists at larger system sizes, with driving amplitudes shifted slightly upward due to the system-size dependence of the yielding transition (\emph{SI Appendix}).

Comparing the two preparation temperatures, both sample sets satisfy the same macroscopic return criterion. However, a consistently larger dissipative response is observed at higher parent temperature, reflected in both the stress-strain curves and the hysteresis-loop area (Fig.~\ref{fig:rpm-loop-both-temps}). Thus, the effect of preparation temperature is not on loop closure itself, but on the distribution of reversible rearrangements throughout the cycle. This motivates a direct investigation of the localized switching units underlying the hierarchical loop structure.

\subsection*{Tile-Switch Hysteron}
\begin{figure}
    \centering
    \includegraphics[width=0.85\columnwidth]{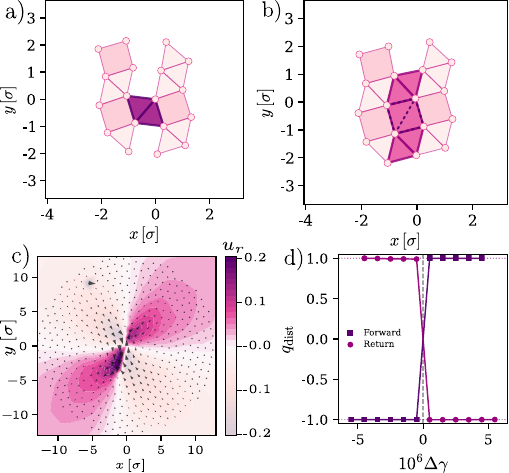}
    \caption{Representative tile-switch hysteron extracted from a refined replay of a matched replica.
    a),~b)~Local square--triangle tiling before and after the forward $\text{A}\to\text{B}$ switch, showing the tiles that rearrange together with the tiles adjacent to them (sharing a vertex). Across the switch, two triangles in state A (dark) are replaced by one square and four  triangles in state B, increasing the local square count from three to four. The dotted lines in panel~b mark the two state-A triangles.
    c)~Eshelby-compatible response around the switching core, showing a quadrupolar displacement field and the  four-lobed shear redistribution in the surrounding quasicrystal. Panels a--c are centered at the $D^{2}_{\min}$-weighted centroid of the switching core.
    d)~Forward and reverse switching traces of the local distance-overlap coordinate, plotted against $10^{6}\Delta\gamma$, where $\Delta\gamma$ is measured relative to the branch-dependent switching midpoint.
    In this shifted representation, both branches exhibit a sharp transition between the A and B state at $\Delta\gamma \approx 0$.
    }
    \label{fig:tile-switch-hysteron}
\end{figure}

To identify the microscopic unit responsible for these reversible excursions, we examine a representative event at a finer strain resolution.
We localize the switching core using the non-affine squared displacement $D^2_{\min}$. Subsequently, we perform a  bond-network reconstruction on the energy-minimized configurations before and after the switch to obtain the corresponding square--triangle tilings (\emph{Materials and Methods}).
Figs.~\ref{fig:tile-switch-hysteron}a-b  show the local tiling immediately before and after an A$\to$B switch on the forward loading branch, occurring between $\gamma=0.004985$ and $0.004986$.
The rearrangement is localized to a connected \emph{switching core} of five particles, defined as the minimal connected set of particles whose bonds rewire between states A and B (\emph{Materials and Methods}).
Figs.~\ref{fig:tile-switch-hysteron}a-b show only the tiles that participate in the rearrangement together with their vertex-sharing neighbors; all other tiles remain unchanged across the switch. All panels are centered at the $D^{2}_{\min}$-weighted centroid of the switching core.
Only a small number of faces change connectivity: the patch evolves from $3$ squares and $8$ triangles to $4$ squares and $12$ triangles, as two triangles in state A are replaced by one square and four triangles in state B.
Fig.~\ref{fig:tile-switch-hysteron}c shows that the switch is embedded in an Eshelby-compatible elastic response (see \emph{SI Appendix}, Fig. S11).
Notably, in the annular region $6\sigma < r < 12\sigma$, which contains $320$ particles, the radial displacement is well described by the Eshelby form
$u_r(r,\theta)\sim \cos[2(\theta-\theta_0)]/r$, consistent with the elastic response of a localized shear transformation~\cite{picardElasticConsequencesSingle2004,puosiTimedependentElasticResponse2014}. The corresponding coarse-grained  shear field exhibits the expected  four-lobed pattern $\epsilon_{xy}(r,\theta)\sim \cos[4(\theta-\theta_0')]/r^2$~\cite{illingStrainPatternSupercooled2016}.
The switch therefore behaves as a localized shear-transformation-like defect embedded in an otherwise elastic quasicrystalline medium.

To directly track the two-state dynamics, Fig.~\ref{fig:tile-switch-hysteron}d plots a local distance-overlap coordinate $q_{\mathrm{dist}}$, defined such that the two template states A and B correspond to $q_{\mathrm{dist}}=-1$ and $q_{\mathrm{dist}}=+1$, respectively.
On the forward branch, $q_{\mathrm{dist}}$ remains near $-1$ up to $\gamma=0.004985$ and jumps to $+1$ at $\gamma=0.004986$.
On the return branch, an equally sharp transition occurs  between $\gamma=-0.048295$ and $-0.048296$.
When the two traces are aligned by their respective switching midpoints, the forward and return transitions collapse onto a single, sharp two-state curve.
Thus, Fig.~\ref{fig:tile-switch-hysteron} identifies the microscopic  memory element as a tile-switch hysteron with a compact core and an Eshelby-like elastic far field. In the \emph{SI Appendix}, we show that  hysteron-like tile-switching units do not possess a unique  local structure  (\emph{SI Appendix},  Fig.S7-S8).
Some involve transitions between square-rich and triangle-rich motifs, while others consist of coordinated rearrangements of several neighboring tiles.
Despite this structural diversity, all tile-switch hysterons  exhibit a compact core and  an Eshelby-like elastic response.

\subsection*{Hierarchical Activity and String-Like Cooperative Motion}
\begin{figure}
    \centering
    \includegraphics[width=0.85\linewidth]{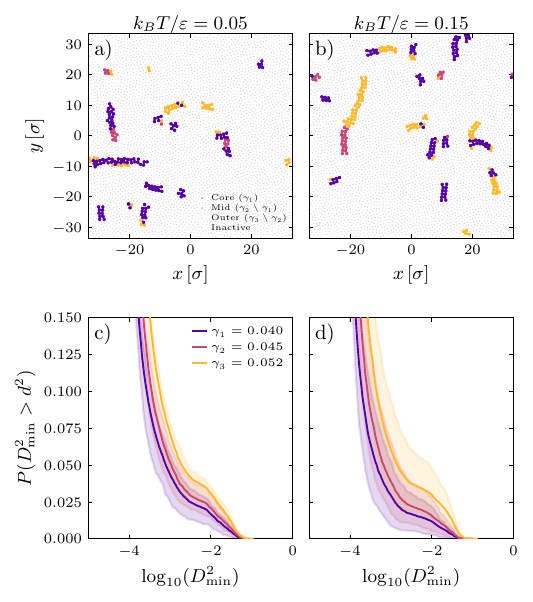}
    \caption{Hierarchical activity maps and complementary cumulative distributions for the nested protocol $\gamma_{3}\to\gamma_{2}\to\gamma_{1}\to\gamma_{2}\to\gamma_{3}$, with $\gamma_{3}=0.052$, $\gamma_{2}=0.045$, $\gamma_{1}=0.04$, and $10$ cycles at each amplitude.
    Panels a and b show representative zero-strain particle maps for low and high parent temperatures, respectively, with particles colored according to the lowest amplitude at which they become active: core ($\gamma_1$), mid ($\gamma_2\setminus\gamma_1$), and outer ($\gamma_3\setminus\gamma_2$); inactive particles are shown in gray.
    Panels c and d show the pooled complementary cumulative distributions $P(D^2_{\min}>d^2)$ for the same three strain amplitudes, aggregated over five  replicas at low and high parent temperatures, respectively. The shaded regions indicate the envelope across replicas.}
    \label{fig:hierarchical-activity}
\end{figure}

To probe the spatial organization of memory, we identify particles that become reversibly active at different strain values and classify them by the lowest value at which they first switch (\emph{Materials and Methods}).
Fig.~\ref{fig:hierarchical-activity}a,b shows that the reversible activity is not organized as an isotropic cloud around a single center.
Instead, active particles form elongated, string-like or worm-like segments that extend over several particle diameters.
The compact switching core identified in the refined hysteron analysis thus represents  only the most localized part of the response; the full reversible event also involves a broader set of particles that move in a correlated, anisotropic manner.
At \(k_{B}T/\epsilon=0.05\), these strings are relatively sparse and thin, and similar  segments are  revisited across all three amplitudes, so the separation between core, mid, and outer activity remains weak.
In contrast, at \(k_{B}T/\epsilon=0.15\) the strings are more abundant and exhibit a  clearer hierarchical structure: as the driving amplitude increases, additional particles are progressively activated along the same elongated pathways, so the cooperative motion expands outward rather than remaining confined to a nearly fixed set of sites.
The complementary cumulative distributions in Fig.~\ref{fig:hierarchical-activity}c,d confirm that this contrast is robust across replicas: at low parent temperature the three amplitude curves lie close together, whereas at high parent temperature each amplitude increase recruits a distinct additional layer of active particles.
The primary difference between the two preparation temperatures is therefore not a qualitatively different local event, but the spatial extent of the cooperative reversible footprint surrounding it: low-temperature samples respond through a restricted set of reversible strings, whereas high-temperature samples support a broader hierarchy of string-like cooperative motion that embeds the same compact switching objects within a richer nested memory structure.
In this picture, the localized tile switch acts as the elementary bistable unit, whereas the macroscopic memory response emerges from the collective activation and elastic coupling of many such units during a shear cycle.

\section*{Discussion}

We have used athermal quasistatic shear to characterize mechanical memory in dodecagonal quasicrystals.
The systems exhibit a yielding transition beyond which the response becomes irreversible, with  persistent shear bands and locally transformed regions of distinct symmetry.
Below yielding, hierarchical cyclic driving generates nested hysteresis loops that return to their pre-excursion limit cycles within the usual steady-state  fluctuations, consistent with loop-return point memory.
This behavior is robust across  preparation temperatures and persists  at larger system sizes.
At the microscopic level, refined replay of  individual reversible events identifies localized phason-like tile rearrangements as the elementary switching units.
These events are sharply bistable in a local distance-overlap coordinate, involve  a compact core-plus-shell structure, and are embedded in an Eshelby-compatible elastic field.
These findings support a picture in which mechanical memory in dodecagonal quasicrystals is carried by tile-switch hysterons, while preparation temperature primarily controls how extensively these reversible units are activated and collectively organized during a cycle. This robustness is not due to  differences in defect content, as both systems are nearly ideal square--triangle tilings with similarly low defect concentrations that remain essentially unchanged   with preparation temperature (\emph{SI Appendix}, Fig.S4). Applying the same protocol  to a defect-free periodic crystalline approximant of the DDQC yields a purely elastic, hysteresis-free response below  yielding, with no nested hysteresis loops. This demonstrates that loop-RPM arises from reversible tile rearrangements enabled by  quasicrystalline disorder rather than from dodecagonal symmetry alone, directly linking phason degrees of freedom of quasicrystals  to the hysteron framework underlying mechanical  memory  (\emph{SI Appendix}).

Perpendicular-space analysis (\emph{SI Appendix}, Fig.S9-S10) further shows that static random-tiling descriptors do not identify the active switching sites, implying that the relevant memory physics is inherently dynamical rather than purely geometric. 
We emphasize that, by combining the non-affine displacement field to localize the switching core of a reversible plastic event with a bond-network reconstruction to obtain  the square--triangle tiling, we can unambiguously identify the particles responsible for each tile-switch hysteron. Such a microscopic identification is considerably more challenging, and perhaps even  impossible, in amorphous solids, where the absence of an underlying structural framework obscures the origin of local particle rearrangements. By systematically increasing the rotational symmetry of quasicrystals \cite{mendozasosaStructuralStudiesLocal2023}, one can continuously interpolate between the dodecagonal quasicrystals studied here and  amorphous glasses. Such systems provide a unique opportunity to uncover how the microscopic mechanisms underlying mechanical memory evolve as structural order is progressively lost,  opening an exciting avenue to explore in future work.
Another natural next step is to construct the transition graph of the observed nested hysteresis loops as formulated by Mungan and Terzi~\cite{munganStructureStateTransition2019a}, or to move beyond the strictly athermal limit and examine whether a small but finite temperature transforms these reversible  cycles into  slowly aging dynamics, analogous to  cyclic aging in  amorphous materials under periodic strain~\cite{shohatAgingAmorphousMaterials2026}.
Extending this framework to other quasicrystal tilings and  symmetries~\cite{dotera2014mosaic} offers another natural direction.

\section*{Materials and Methods}

\subsection*{Model and sample preparation}
We model a 2D DDQC using the continuous interaction potential $U_{\mathrm{QC}}(r) / \epsilon = {\left( \sigma/r \right)}^{14} + (1 - \tanh{\left[ k(r - \delta)\right]})/2$, where $\epsilon$ and $\sigma$ define the units of energy and length, respectively, $k\sigma=10$ controls the steepness of the shoulder, and $\delta=1.35\sigma$ sets the interaction range. These parameters are chosen to stabilize the DDQC phase~\cite{kryuchkov2018complex,padillaPhaseBehaviorTwodimensional2020,coliInverseDesignSoft2021}. The potential is smoothly truncated with the XPLOR switching function as implemented in HOOMD-blue~\cite{anderson2020hoomd},
\begin{equation}
    S(r) =
    \begin{cases}
        1 & r \le r_{\mathrm{on}} \\[4pt]
        \dfrac{(r_c^2 - r^2)^2 (r_c^2 + 2r^2 - 3r_{\mathrm{on}}^2)}{(r_c^2 - r_{\mathrm{on}}^2)^3} & r_{\mathrm{on}} < r < r_c \\[4pt]
        0 & r \ge r_c \, ,
    \end{cases}
    \label{eq:xplor}
\end{equation}
applied over the range $r_{\mathrm{on}}=1.9\sigma$ to the cutoff distance $r_c=2.0\sigma$, such that both the potential and its derivative vanish smoothly at $r_c$. The resulting interaction is given by $U(r)=U_{\mathrm{QC}}(r)\,S(r)$.

DDQC samples containing $N=4096$ or $N=16384$ particles are prepared via $NVT$ molecular dynamics simulations in LAMMPS~\cite{THOMPSON2022108171,plimpton_2024_10806852}, with the interaction potential and corresponding forces implemented as native pair styles rather than through tabulated inputs. We initialize a disordered liquid of particles of mass $m$ in a square simulation box at number density $\rho\sigma^2=0.94$, using periodic boundary conditions. The system is first equilibrated at $k_BT/\epsilon=1$ for $10^6$ timesteps, then slowly annealed to the target parent temperature over $10^7$ timesteps, and subsequently equilibrated at that temperature for at least $10^8$ timesteps using the Bussi \emph{et al.}\ stochastic velocity-rescaling thermostat~\cite{bussiCanonicalSamplingVelocity2007}, with a coupling time of $0.1\tau$. Particle positions are integrated using a velocity-Verlet algorithm with a timestep $\Delta t=0.005\tau$, where $\tau=\sigma\sqrt{m/\epsilon}$ sets the unit of time. We prepare samples at two parent temperatures, $k_BT/\epsilon=0.05$ and $0.15$, to test whether loop-RPM is sensitive to preparation history; the two sample sets are self-assembled independently at their respective preparation temperatures and therefore represent distinct microstructures. In both cases, the final equilibrated configurations are energy-minimized using the FIRE algorithm~\cite{bitzek2006structural,restrepo2023abc} until the infinity norm of the force vector, $\|\mathbf{F}\|_\infty=\max_{i,\alpha}|F_{i,\alpha}|$, falls below $10^{-10}\,\epsilon/\sigma$, at fixed density $\rho\sigma^2=0.94$ (see \emph{SI Appendix}).

\subsection*{Athermal quasistatic shear}
The resulting configurations are subjected to an athermal quasistatic shear (AQS) protocol, implemented as an iterative two-step procedure. In the first step, an affine shear deformation is applied to all particle positions through the coordinate transformation $x' \rightarrow x + \Delta \gamma\, y$, $y'\rightarrow y$, corresponding to a simple shear strain increment $\Delta\gamma=10^{-4}$: the $x$-coordinate of each particle $i$ is shifted by $\Delta\gamma\,y_i$. The simulation box is deformed accordingly, and Lees--Edwards periodic boundary conditions are applied. In the second step, the deformed configuration is energy-minimized using the FIRE algorithm to the same force tolerance $\|\mathbf{F}\|_\infty<10^{-10}\,\epsilon/\sigma$, ensuring that the system remains at a local energy minimum before the next strain increment is applied. The two-step procedure is repeated until the strain reaches the target maximum strain amplitude $\gamma_{\mathrm{max}}$. For cyclic AQS, the sign of $\Delta\gamma$ is reversed whenever $\gamma=\pm\gamma_{\mathrm{max}}$, so that a complete shear cycle follows $0\to\gamma_{\mathrm{max}}\to 0\to -\gamma_{\mathrm{max}}\to 0$, corresponding to a total accumulated absolute strain of $4\gamma_{\mathrm{max}}$ per cycle. Steady-state limit cycles are identified by monitoring the energy per particle $U/N$ and the loop area $\Sigma=\oint\sigma_{xy}\,d\gamma$ at the end of each cycle; convergence is declared when both quantities remain unchanged, within numerical precision, between successive cycles. In practice, this steady state is reached within $N_c=2$--$10$ cycles depending on $\gamma_{\mathrm{max}}$ and the parent temperature of the sample.

\subsection*{Bond-network reconstruction}
We characterize the local structure using  two complementary graph representations constructed from the same particle positions. To obtain the square--triangle tiling, we first connect all particle pairs separated by less than $1.35\sigma$. Each bond is then assigned to the nearest of the twelve dodecagonal bond directions (separated by $30^\circ$), provided its orientation differs by no more than $7.5^\circ$ from the corresponding ideal direction~\cite{oxborrowRandomSquaretriangleTilings1993,imperor-clercSquaretriangleTilingsInfinite2021}. The enclosed faces are subsequently classified as triangles, squares, or defects, with the latter including shield- and egg-shaped tiles~\cite{ulugol2026vacancydefectssquaretriangletilings}.  This construction yields the square--triangle tiling shown in the tile panels.  Separately, we compute the Delaunay triangulation of each  configuration using  \texttt{freud}~\cite{freud2020}. Because the Delaunay triangulation is uniquely defined, it provides an unambiguous neighbor set $\mathcal{N}(i)$ for evaluating $D^2_{\min}$.  Crucially, tile-switch events are identified from the rewiring of the Delaunay network---that is, from the edges that are created or removed between the pre- and post-switch configurations---rather than from the appearance or disappearance of reconstructed tiles. These bond changes directly localize the switching core. The square--triangle tiling is subsequently reconstructed for both configurations to visualize the associated tile rearrangement. 

\subsection*{Non-affine displacement and tile-switch identification}
To localize the switching core of a reversible plastic event, we use the non-affine squared displacement~\cite{falk1998dynamics},
\begin{equation}
    D^2_{\min,i} = \frac{1}{N_i} \min_{\mathbf{J}_i} \sum_{j \in \mathcal{N}(i)} \left| \Delta \mathbf{r}_{ij}^{\,\text{post}} - \mathbf{J}_i \, \Delta \mathbf{r}_{ij}^{\,\text{pre}} \right|^2 \, ,
    \label{eq:d2min}
\end{equation}
where $\Delta \mathbf{r}_{ij} = \mathbf{r}_j - \mathbf{r}_i$ denotes the relative position  of neighbor $j$ with respect to particle $i$, $\mathcal{N}(i)$ represents the set of $N_i$ Delaunay neighbors of particle $i$ introduced above, and $\mathbf{J}_i$ is the best-fit local deformation gradient tensor. The superscripts ``pre'' and ``post'' refer to the energy-minimized configurations immediately before and after the localized switching event, respectively. Subsequently, we perform a bond-network reconstruction on both energy-minimized configurations to obtain the corresponding square--triangle tilings. 

We resolve an individual tile-switch hysteron by replaying the trajectory near the switching strain with progressively smaller shear increments, until the transition is captured between two consecutive energy-minimized states, A and B. Since these rewired bonds can be distributed over multiple disjoint sites, we keep only the connected cluster associated with the largest peak in the non-affine displacement $D^{2}_{\min}$~[\eqref{eq:d2min}], which isolates a single localized switching event. The \emph{switching core} is then defined as the smallest connected set of particles that contains at least one endpoint of every rewired bond in that cluster. For the representative event discussed in the main text, this core consists of five particles.

The displacement- and strain-field analysis of the switching core is performed in a coordinate frame centered on the $D^{2}_{\min}$-weighted centroid of the switching core; the same origin is used for the tiling panels. This centroid lies within approximately $0.4\sigma$ of the geometric center of the rearranging tiles, so the chosen origin coincides closely with the structural center of the switch. The Eshelby fit (see \emph{SI Appendix}) is performed over the annular region $6\sigma<r<12\sigma$, measured from this origin.

\subsection*{Hierarchical-activity classification}
To examine how the reversible response is organized in real space throughout the nested loading protocol, we analyze the peak non-affine activity at the maximum strain of each branch in the sequence $\gamma_3\to\gamma_2\to\gamma_1\to\gamma_2\to\gamma_3$, with $\gamma_3=0.052$, $\gamma_2=0.045$, and $\gamma_1=0.04$. For each replica and each branch, we recompute the affine-corrected peak $D^2_{\min}$ field after subtracting the reversible affine shear, using the same local non-affine measure as in ~\eqref{eq:d2min}. To classify particles as active or inactive, we employ a data-driven threshold defined by the gap in the pooled $D^2_{\min}$  distribution at $\gamma_3$, yielding $D^2_{\min,\mathrm{th}}=5.05\times10^{-3}$ for the low-temperature ensemble and $5.99\times10^{-3}$ for the high-temperature ensemble. Each particle is then assigned to the lowest strain amplitude at which it first becomes active. We classify particles as  \emph{core} if they are already active  at $\gamma_1$, as  \emph{mid} if they first become active at  $\gamma_2$ but not at $\gamma_1$, and as \emph{outer} if they  become active only at $\gamma_3$. If the microscopic response reflects the same hierarchy as the macroscopic nested hysteresis loops, the corresponding active sets should themselves form  a nested hierarchy, $S(\gamma_1)\subseteq S(\gamma_2)\subseteq S(\gamma_3)$.

\section*{Data Availability}
The simulation, analysis, and plotting code, together with the processed data required to reproduce the figures, are publicly available on Zenodo at \href{https://doi.org/10.5281/zenodo.21281566}{https://doi.org/10.5281/zenodo.21281566}.

\begin{acknowledgments}
E.A.B.-M.\ thanks Andr\'e Matias and Nex Stuhlm\"uller for many helpful discussions. M.D.\ and E.A.B.-M.\ acknowledge funding from the European Research Council (ERC) under the European Union's Horizon 2020 research and innovation programme (Grant agreement No.\ ERC-2019-ADG 884902 SoftML).
\end{acknowledgments}

\bibliography{bibliography}

\end{document}


\title{Supporting Information for:\\ How Quasicrystals Remember: Hierarchical Memory Under Cyclic Shear}

\author{Edwin A. Bedolla-Montiel}
 \email{e.a.bedollamontiel@uu.nl}
\affiliation{Soft Condensed Matter \& Biophysics, Debye Institute for Nanomaterials Science, Utrecht University, Princetonplein 1, 3584 CC Utrecht, Netherlands.}

\author{Marjolein Dijkstra}
 \email{m.dijkstra@uu.nl}
\affiliation{Soft Condensed Matter \& Biophysics, Debye Institute for Nanomaterials Science, Utrecht University, Princetonplein 1, 3584 CC Utrecht, Netherlands.}

\date{\today}
\maketitle

\section*{Simulation Protocol and Sample Characterization}\label{ssec:simdetails}

The complete simulation and shearing protocol---including the interaction potential and its XPLOR truncation, the $NVT$ molecular-dynamics preparation at the two parent temperatures, the FIRE energy minimization, the athermal quasistatic shear (AQS) protocol, and the reconstruction of the square--triangle bond network---is described in the \emph{Materials and Methods} of the main text. Here, we present   supplementary figures that characterize the resulting samples and illustrate the protocol.

\begin{figure}
    \centering
    \includegraphics[width=\textwidth]{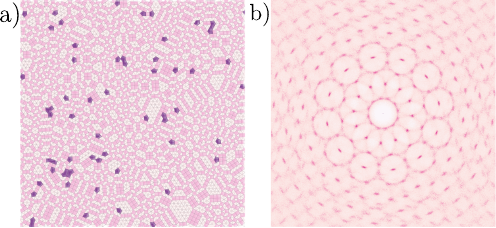}
    \caption{Quasicrystal sample prepared at \(k_{B}T/\epsilon=0.05\) and subsequently energy-minimized. Panel a) shows the reconstructed square--triangle bond network, obtained as described under \emph{Bond-network reconstruction} in the \emph{Materials and Methods} of the main text; faces that do not satisfy the tiling rules are labeled as defects, even when they appear visually similar to distorted or nearly regular squares or triangles. Panel b) shows the corresponding diffraction  intensity map computed from the same minimized configuration.}
    \label{fig:quasicrystal sample}
\end{figure}

\begin{figure}
    \centering
    \includegraphics[width=\textwidth]{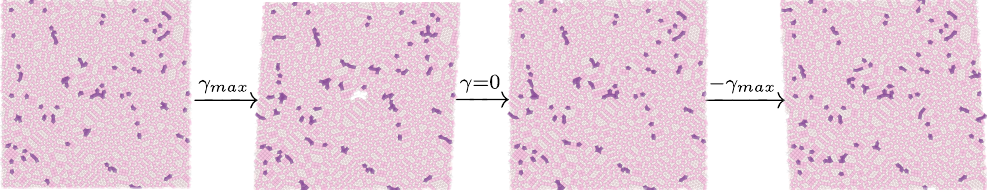}
    \caption{Athermal quasistatic shear (AQS) cycle visualized for a representative \(N=4096\) dodecagonal quasicrystal configuration at the maximum strain amplitude \(\gamma_{\max}=0.052\) used for the return point memory loops in the main text. The panels show the reconstructed square-triangle network  at successive stages of the cycle, \(\gamma=0 \rightarrow +\gamma_{\max} \rightarrow 0 \rightarrow -\gamma_{\max}\rightarrow 0\), with the final step returning the system to zero strain. In all panels, closed faces that do not satisfy the square-triangle tiling rules (see \emph{Bond-network reconstruction} in the \emph{Materials and Methods} of the main text) are classified as defects and highlighted in dark purple, while white regions indicate  locations where no closed tiles could be reconstructed. The cyclic deformation remains essentially reversible at this sub-yield amplitude. A small white patch of unreconstructed tiles appears near the center at the peak strain \(+\gamma_{\max}\), then heals as the strain returns to zero and reverses, so the tiling recovers its original square-triangle order by the end of the cycle. The order lost under load is thus fully restored on unloading, with no net accumulation over the loop, mirroring the macroscopic loop-return point memory reported in the main text.}
    \label{fig:aqs_cycle}
\end{figure}

\begin{figure}
    \centering
    \includegraphics[width=\textwidth]{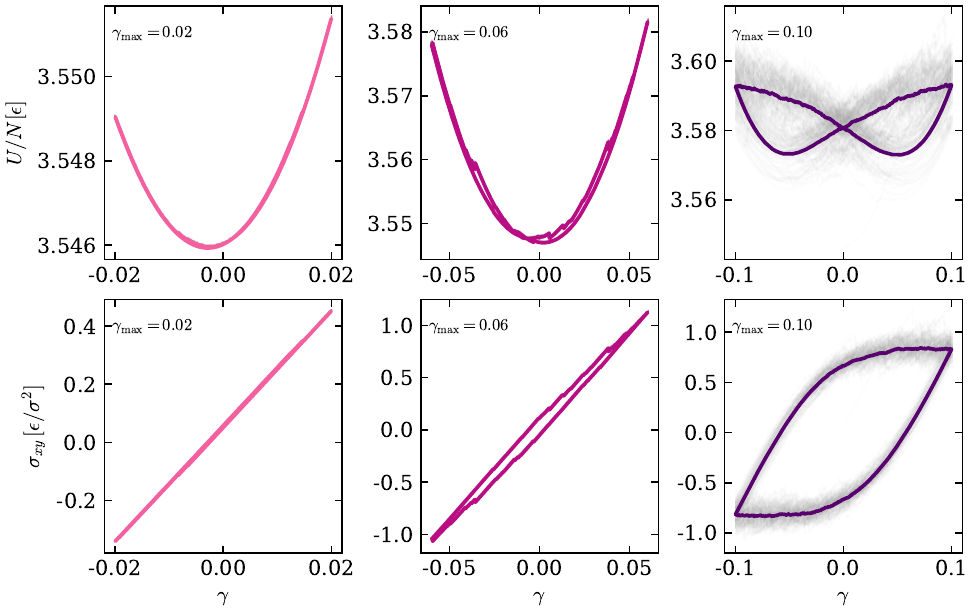}
    \caption{Mechanical response under AQS cyclic shear for a quasicrystal sample of \(N=4096\) particles. Columns correspond to  maximum strain amplitudes \(\gamma_{\max}=0.02\), \(0.06\), and \(0.10\), using a strain resolution of \(\Delta \gamma = 10^{-4}\) and evolved for \(100\) cycles. The top row shows the energy per particle \(U/N\), and the bottom row shows the shear stress \(\sigma_{xy}\), both as functions of strain \(\gamma\). Thin gray curves represent individual cycles from a single replica, while colored curves denote the pointwise cycle-averaged response. As \(\gamma_{\max}\) increases,  the response evolves from nearly reversible elastic behavior to  pronounced hysteresis. The slight asymmetry observed in the \(\gamma_{\max}=0.02\) energy curve is consistent with residual pre-strain or pre-stress in the quasicrystal, which can arise when an inherently aperiodic quasicrystalline structure is embedded in a finite periodic simulation box.}
    \label{fig:quasicrystal response}
\end{figure}

Fig.~\ref{fig:quasicrystal sample} shows a representative energy-minimized quasicrystal sample prepared at \(k_{B}T/\epsilon=0.05\), together with its corresponding diffraction pattern.
Fig.~\ref{fig:aqs_cycle} visualizes a full AQS cycle at the sub-yield strain amplitude \(\gamma_{\max}=0.052\) used for the return point memory loops in the main text, illustrating the reversibility of the tiling throughout the cycle.
Steady-state limit cycles are identified by monitoring the energy per particle \(U/N\) and the loop area \(\Sigma=\oint\sigma_{xy}\,d\gamma\) at the end of each cycle (see \emph{Materials and Methods}). In practice, the steady state is reached within \(2\)--\(10\) cycles, depending on the strain amplitude \(\gamma_{\mathrm{max}}\) and the parent temperature. Fig.~\ref{fig:quasicrystal response} shows the resulting mechanical response as the strain amplitude increases from \(\gamma_{\max}=0.02\) to \(0.10\).

\section{Effect of Preparation Temperature on Defect Concentration}\label{ssec:defects}

To compare the two sample sets at preparation temperature $k_BT/\epsilon=0.05$ and 0.15, we reconstruct the square--triangle tiling (\emph{Bond-network reconstruction} in \emph{Materials and Methods}) of each zero-strain minimized configuration and quantify its defect content following the procedure of Ref.~\cite{ulugol2026vacancydefectssquaretriangletilings}. Since an ideal dodecagonal tiling consists exclusively of  regular squares and equilateral triangles, we define the defect content as the area fraction of all tiles that are neither squares nor equilateral triangles, including the so-called shield- and egg-shaped defects as well as any other irregular tiles.

\begin{figure}
    \centering
    \includegraphics[scale=1.5]{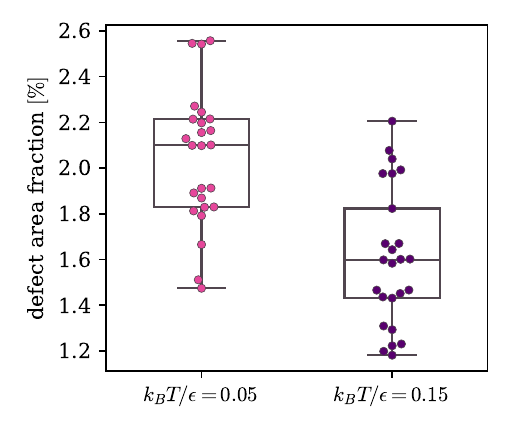}
    \caption{Defect content of the two sample sets. Area fraction of all irregular tiles, defined as all tiles that are neither regular squares nor triangles, in the reconstructed square--triangle tiling of each zero-strain minimized configuration ($N=4096$, $25$ replicas per temperature). Boxes indicate the median and interquartile range, while points are individual replicas. The defect content is low at both preparation temperatures and is, if anything, slightly lower in the higher-temperature ensemble.}
    \label{fig:defect-area-fraction}
\end{figure}

Both sets remain close to ideal square--triangle tilings: averaged over  $25$  replicas, squares and triangles occupy equal total areas  within $1\%$, yielding a tile-number ratio of $N_\triangle/N_\square = A_\square/A_\triangle = 2.26$, where $A_\square$  and $A_\triangle$ denote the mean areas of square and triangle tiles, respectively, at both temperatures.
The overall defect content is low in both cases (Fig.~\ref{fig:defect-area-fraction}), approximately $\sim\!2.0\times10^{-2}$ at $k_BT/\epsilon=0.05$ and $\sim\!1.6\times10^{-2}$ at $k_BT/\epsilon=0.15$. The shield and egg defects constitute only a small fraction of this total (approximately $\sim\!8\times10^{-4}$ and $\sim\!4\times10^{-4}$, respectively). The defect content  does not increase with preparation temperature; if anything, it is slightly lower at higher temperature for both  irregular and shield/egg tiles.

This trend is opposite to equilibrium expectations, where the  free-energy cost of shield and egg defects decreases with temperature and their concentration should therefore increase~\cite{ulugol2026vacancydefectssquaretriangletilings}.  We thus attribute the enhanced activity and cooperative motion at higher temperature to kinetic effects. Although both sets are annealed for the same number of time steps prior to energy minimization, activated tile rearrangements that relax the tiling are significantly slower at $k_BT/\epsilon=0.05$. As a result, these samples likely equilibrate their defect concentration more slowly within the simulated time and retain a higher fraction of irregular tiles.
Nevertheless,  the static defect density remains low and nearly identical across both temperatures, indicating that the broader hysteresis-loop excursions cannot be explained by differences in  defect concentration. Instead, the  temperature dependence must arise from differences in the dynamical organization of the localized reversible tile-switch rearrangements.

\section{Shear Bands and Structural Evolution}\label{ssec:shear}

To characterize the structural evolution of the DDQC in the plastic regime, we shear samples of $N=16384$ particles at $\gamma_{\max}=0.1>\gamma_y$ for $N_c=100$ cycles without waiting for a steady state, as the aim is to track the progressive structural transformation of the material.
Local structural order is quantified using the $l$-fold bond order parameter
\begin{equation}
    q_{l}(i) = \frac{1}{\mathcal{N}(i)} \sum_{j \in \mathcal{N}(i)} e^{il\theta_{ij}} \, ,
    \label{eq:si_bop}
\end{equation}
where $\mathcal{N}(i)$ denotes the set of nearest neighbors of particle $i$, identified via a cutoff  radius of $r_c=1.35\sigma$, consistent with the shoulder length of the interaction potential, and $\theta_{ij}$ is the bond angle relative to a fixed reference axis.
The magnitude $|q_l(i)|$ ranges from $0$ (no $l$-fold orientational order) to $1$ (perfect $l$-fold symmetry). We  compute this quantity  for $l=4,6,12$ to distinguish square ($l=4$), hexagonal ($l=6$), and dodecagonal ($l=12$) local environments, respectively.
\begin{figure}
    \centering
    \includegraphics[width=\textwidth]{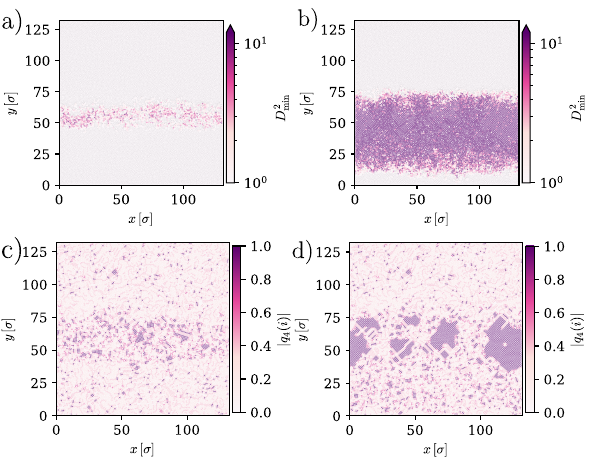}
    \caption{Panels a and b depict the non-affine displacement $D^{2}_{\min}$, defined in the \emph{Materials and Methods} of the main text, for a DDQC sample of $N=16384$ particles prepared at \(k_{B}T/\epsilon=0.05\) and sheared at $\gamma_{\max}=0.1$. Particles are colored according to the magnitude of $D^{2}_{\min}$ after a) $N_c=5$ cycles and b) $N_c=100$ cycles, revealing the progressive accumulation of shear localization in the central region. Panels c and d show corresponding spatial maps of local $l=4$ bond orientational order, quantified by  bond order parameter \(q_{4}(i)\)~[Eq.~\eqref{eq:si_bop}]. Regions with large  $|q_4(i)|$ form near the center after $N_{c}=5$ cycles and subsequently consolidate into a persistent shear band after $N_{c}=100$ cycles.}
    \label{fig:spatial-maps}
\end{figure}

Spatial maps of $|q_l(i)|$  for $l=4$ are shown in Fig.~\ref{fig:spatial-maps} for the system after  $N_c=5$ (left  column) and $N_c=100$  (right column) cycles, together with corresponding maps of the non-affine displacement $D^{2}_{\min}$ (\emph{Materials and Methods} of the main text).
After $N_c=5$ cycles, a region of enhanced $4$-fold symmetry already appears near the center of the sample, while $12$-fold symmetry is locally suppressed in the same region (not shown). This indicates  the emergence of locally transformed patches within the quasicrystal.
The $6$-fold order parameter shows no significant spatial structure at this stage (not shown).
After $N_c=100$ cycles, the central region is dominated by a consolidated band of enhanced $4$-fold and suppressed $12$-fold order (high $|q_4|$, low $|q_{12}|$).
These observations are consistent with results for amorphous solids~\cite{bhaumik2022yielding} and recent quasicrystal simulations~\cite{maireSelfassemblyQuasicrystalsCyclic2026}, which showed that such transformed regions can persist indefinitely in the plastic regime.
The structural transition away from quasicrystalline order thus provides the microscopic origin of the energy dissipation and hysteresis loops observed above $\gamma_y$ in the main text.

\begin{figure}
    \centering
    \includegraphics[scale=1.5]{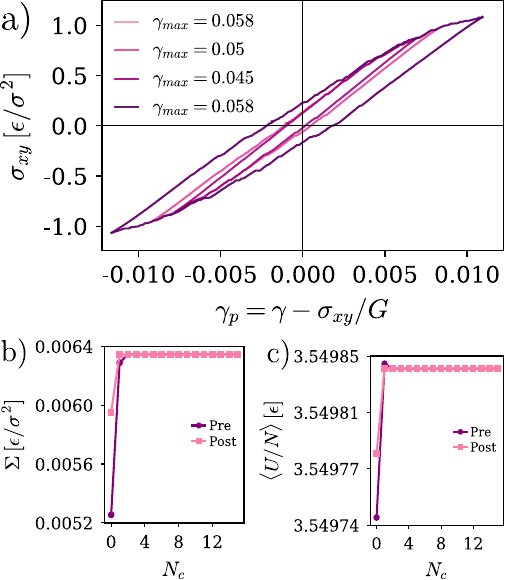}
    \caption{Hierarchical cyclic-shear response of a 2D dodecagonal quasicrystal with  $N=16384$ particles prepared at parent temperature $k_B T/\epsilon=0.15$. The nested protocol follows $\gamma_3\to\gamma_2\to\gamma_1\to\gamma_2\to\gamma_3$, with $\gamma_3=0.058$, $\gamma_2=0.05$, $\gamma_1=0.045$, and $N_c=15$ cycles at each amplitude. a) shear stress $\sigma_{xy}$ as a function of plastic strain $\gamma_p=\gamma-\sigma_{xy}/G$, where $G$ is obtained from a fit to the initial loading branch. b)  Hysteresis-loop area $\Sigma$ and c) energy per particle $U/ N$ as a function of cycle number $N_c$.}
    \label{fig:rpm-loop-both-temps-largeN}
\end{figure}

\section{Finite-Size Effects}\label{ssec:finite-size}

To verify that our results on loop-return point memory are persistent at larger system sizes, we apply the hierarchical protocol $\gamma_{3} \to \gamma_{2} \to \gamma_{1} \to \gamma_{2} \to \gamma_{3}$, with $\gamma_{3} = 0.058$, $\gamma_{2} = 0.05$, $\gamma_{1} = 0.045$, and $N_{c} = 10$ cycles at each amplitude, to a system of $N=16384$ particles prepared at density $\rho\sigma^{2}=0.94$ and temperature $k_{B}T/\epsilon=0.15$.
As shown in Fig.~\ref{fig:rpm-loop-both-temps-largeN}, the system exhibits  the same nested hysteretic response, including well-defined  subloops during the excursions.

Applying the RPM criterion from the main text,  we confirm that all $n=10$ replicas  return to their  previously visited states within numerical precision.
Note that the strain amplitudes are shifted upward relative to the $N=4096$ protocol, since increasing the system size raises the yielding transition to $\gamma_y\approx 0.07$, thereby shifting the strain region in which RPM is observed.

\section{Microscopic Tile-Switching Events in  Quasicrystalline Samples}\label{ssec:tile-switching}
\begin{figure}
  \centering
  \includegraphics[width=\textwidth]{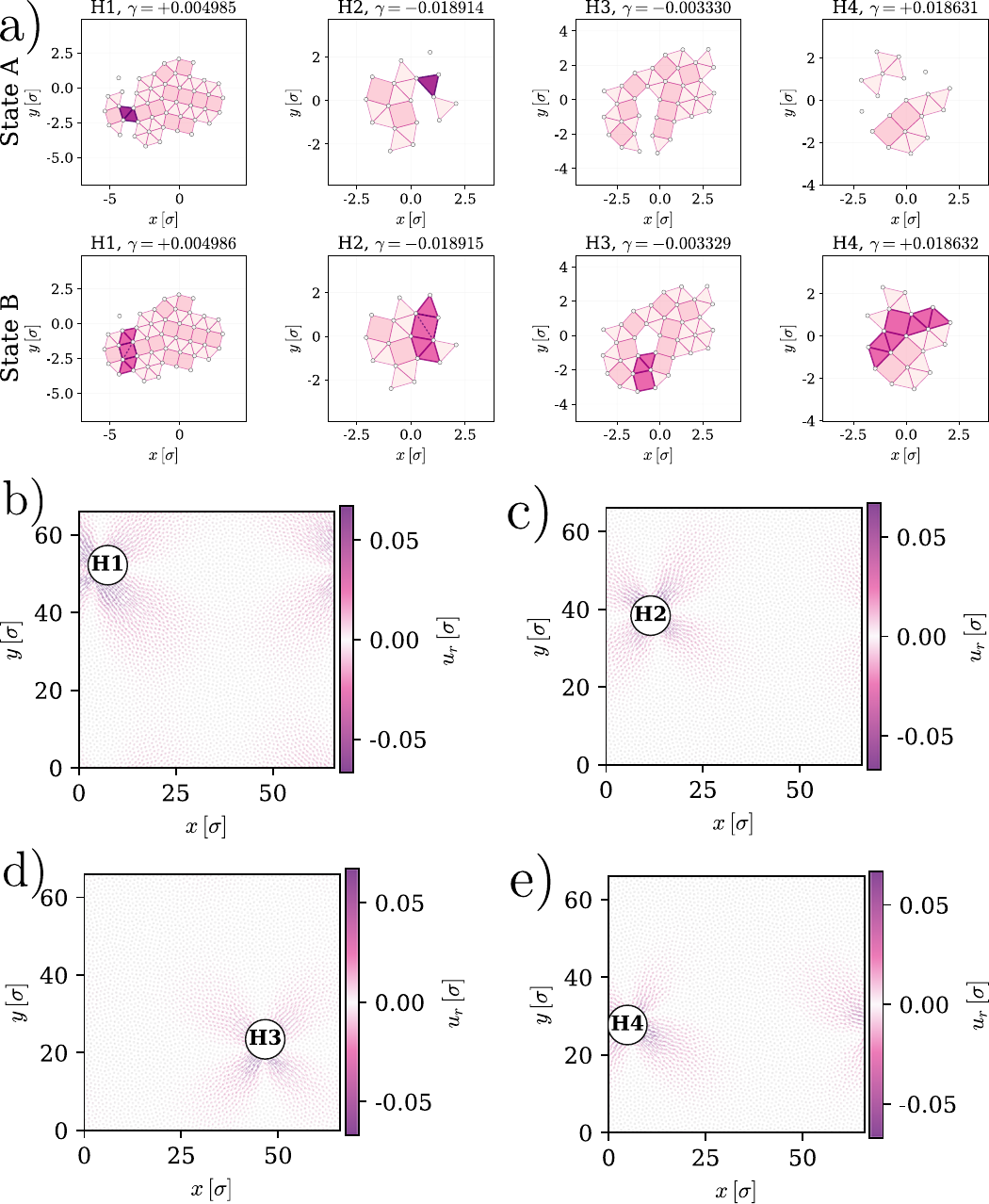}
  \caption{
  Representative localized tile-switching units in a low-temperature quasicrystalline sample. The example is taken from a single replica with $N=4096$, $\rho \sigma^{2}=0.94$, preparation temperature $k_{B}T/\epsilon=0.05$, and a limit cycle at $\gamma_{\max}=0.052$. a) Local square-triangle tiling for four localized hysteron-like switches, H1--H4. For each unit, states A and B correspond to neighboring minimized configurations across the local switching event; darkened tiles and edges mark the associated topological changes. b--e) Full-system displacement fields corresponding to the  switching events H1, H2, H3, and H4, respectively. Gray points denote particle positions before the switch, arrows indicate minimum-image displacement vectors, and color encodes the unsigned radial displacement component $u_r$, shown on a symmetric scale centered about zero. The H1--H4 labels mark the localized tile-switching regions and do not necessarily coincide with the  apparent center of the largest particle displacements. Arrow lengths are visually rescaled for clarity.}
  \label{fig:supp-hysteron-units}
\end{figure}

Fig.~\ref{fig:supp-hysteron-units} presents a representative set of localized hysteron-like switching units identified within a single low-temperature quasicrystalline sample.
The example is taken from one replica containing $N=4096$ particles at number density $\rho\sigma^{2}=0.94$, preparation temperature $k_{B}T/\epsilon=0.05$, and driven into a limit cycle with maximum strain amplitude $\gamma_{\max}=0.052$, well within the return point memory regime discussed in the main text.
The four units shown, labeled H1--H4, are not intended to represent an ensemble average, rather they provide  concrete microscopic examples of how localized tile rearrangements couple to long-ranged elastic displacement fields. The tiling panels show the local square-triangle tiling reconstructed before and after the same switching event.

Light-colored tiles are common to both local states, whereas the darker highlighted tiles and edges indicate the local topological changes between state A and B.
Particle markers  are kept neutral in the tiling panels so that the figure emphasizes the tile rearrangement itself, rather than the specific set of particles used to identify the switching core.

It is also important to note that the switching units generally involve more than a single tile.
This is evident in all four examples H1--H4 in Fig.~\ref{fig:supp-hysteron-units}a.
In some cases, such as H2, the local structure in state A is close to forming a square- or triangle-like tile, but the corresponding configuration is not yet stabilized in the energy-minimized configuration.
The imposed quasistatic shear then drives the system into a nearby minimum in which the local square-triangle tiling changes discontinuously.
Thus, the hysteron-like response is  not  a single-particle rearrangement, but rather a localized topological tile switch embedded within the quasiperiodic tiling network.

For each unit H1--H4, we identify two neighboring minimized configurations along the refined replay trajectory, denoted state A and B, on either side of the first-order  transition in the local distance-overlap order parameter.
The particle displacement field is then computed as
\begin{equation}
    \Delta \mathbf{r}_j
    =
    \mathbf{r}^{B}_j - \mathbf{r}^{A}_j \, ,
    \label{eq:displacement-field}
\end{equation}
using the minimum-image convention within the periodic simulation cell.
The displacement panels display this vector field for all $N=4096$ particles.
Gray dots indicate particle positions in state A, while arrows represent the corresponding displacement
vectors.
Arrow lengths are uniformly rescaled for visual clarity and therefore reflect relative rather than absolute displacement magnitudes.

To visualize the angular structure of the response, each particle is colored according to the radial component of its displacement,
\begin{equation}
    u_r
    =
    \Delta \mathbf{r}_j \cdot \hat{\mathbf{r}}_j \, ,
    \label{eq:radial-displacement-field}
\end{equation}
where $\hat{\mathbf{r}}_j$ is the unit vector pointing from the localized switching region to particle $j$, evaluated using periodic boundary conditions.
The color scale is chosen to be symmetric about zero, such that inward and outward radial displacements of equal magnitude are assigned equal visual weight.

The examples in Fig.~\ref{fig:supp-hysteron-units}a illustrate that hysteron-like tile-switching units do not share a unique  local structure.
Some involve transitions between square-rich and triangle-rich motifs, while others consist of coordinated rearrangements of several neighboring tiles.
Despite this structural diversity, the displacement fields shown in Fig.~\ref{fig:supp-hysteron-units}b--e all exhibit a long-ranged quadrupolar pattern consistent with an Eshelby-like elastic response, in agreement with the mechanism discussed in the main text.

\begin{figure}
  \centering
  \includegraphics[width=0.85\textwidth]{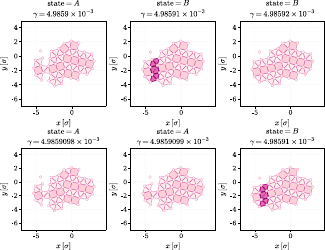}
  \caption{Resolution test for the localized tile-switching mechanism. The top row shows unit H2 resolved with a strain step $\Delta\gamma = 10^{-8}$, while the bottom row shows the same unit resolved with $\Delta\gamma = 10^{-10}$. Each row shows three consecutive energy-minimized configurations from the refined trajectory. The state label indicates whether the local distance-overlap order parameter assigns the configuration to state A or state B, and the corresponding value of $\gamma$ gives the applied shear strain at that energy-minimized configuration. The highlighted tiles and edges mark the local topological change associated with the switching event. The comparison shows that increasing the strain resolution localizes the switch to a narrower strain interval without altering the underlying tile-switching mechanism.}
  \label{fig:supp-hysteron-resolution}
\end{figure}

We further test whether the resolution used in the shear cycle affects the identified tile-switching units by revisiting unit H2 and replaying the trajectory around the relevant strain interval at two different resolutions, namely $\Delta\gamma = 10^{-8}$ and $\Delta\gamma = 10^{-10}$.
Fig.~\ref{fig:supp-hysteron-resolution} shows the configurations immediately before and after the tile-switching event at resolution $\Delta\gamma = 10^{-8}$ in the top panels, while the bottom panels display the same event resolved at $\Delta\gamma = 10^{-10}$.
We find that the tile-switching event is not stepwise in nature and does not exhibit a hierarchy of cascading rearrangements, as would be expected for an avalanche in which a plastic event triggers subsequent events,  reported for amorphous solids~\cite{bhaumik2022yielding}.
Instead, increasing the resolution merely localizes the tile-switching event to a  narrower strain interval, without altering the underlying mechanism or its reversibility.

\section{Parallel and Perpendicular Space Analysis}\label{ssec:parallel}

To investigate whether reversible switching sites can be predicted from the static tiling geometry alone, we perform a lifted-space analysis on the full protocol-matched dataset, consisting of $25$ low-temperature and $25$ high-temperature replicas.
For each zero-strain reference configuration, we reconstruct the square-triangle bond network, removed the immediate neighborhoods of tiling defects, and retain the largest connected region for  which a consistent lift can be constructed.
The lifting procedure assigns to each vertex in this defect-filtered core  an integer coordinate $\mathbf{n}_i=(n_{i0},n_{i1},n_{i2},n_{i3})\in\mathbb{Z}^4$ in the four-dimensional hyperlattice associated with  the dodecagonal quasicrystal (DDQC).
We employ the standard DDQC embedding,  whose projections onto physical (parallel) space and internal (perpendicular) space are given by~\cite{oxborrowRandomSquaretriangleTilings1993}
\begin{equation}
  \mathbf{e}^{\parallel}_{m}
  =
  \left(\cos\frac{m\pi}{6},\,\sin\frac{m\pi}{6}\right),
  \qquad
  \mathbf{e}^{\perp}_{m}
  =
  \left(\cos\frac{7m\pi}{6},\,\sin\frac{7m\pi}{6}\right),
  \qquad m=0,\ldots,3 .
\end{equation}
Let $\{\mathbf{g}_m\}_{m=0}^{3}$ denote the standard basis vectors of the integer hyperlattice $\mathbb{Z}^4$.
In this representation, the square-triangle tiling is generated by six positive nearest-neighbor steps
\begin{equation}
  \mathbf{g}_0,\quad
  \mathbf{g}_1,\quad
  \mathbf{g}_2,\quad
  \mathbf{g}_3,\quad
  \mathbf{g}_2-\mathbf{g}_0,\quad
  \mathbf{g}_3-\mathbf{g}_1 \, ,
\end{equation}
together with their negatives.
After projection using the parallel-space basis $\mathbf{e}^{\parallel}_{m}$, these twelve lattice steps map onto the twelve allowed bond directions of the DDQC, which are uniformly spaced  by $30^\circ$.

Up to a  replica-dependent scale factor  $a_r$ for replica $r$, the lifted coordinates are  projected into parallel and perpendicular space according to
\begin{equation}
  \mathbf{x}^{\parallel}_{i}
  =
  a_r \sum_{m=0}^{3} n_{im}\mathbf{e}^{\parallel}_{m},
  \qquad
  \mathbf{x}^{\perp}_{i}
  =
  a_r \left[
  \sum_{m=0}^{3} n_{i \, m}\mathbf{e}^{\perp}_{m}
  -
  \left\langle
  \sum_{m=0}^{3} n_{j\, m}\mathbf{e}^{\perp}_{m}
  \right\rangle_{j\in\mathcal{V}_r}
  \right],
  \qquad i\in\mathcal{V}_r \, .
\end{equation}
Here, $\mathcal{V}_r$ denotes the defect-filtered connected core. The subtraction of the mean perpendicular-space coordinate removes the arbitrary origin associated with  the open-boundary lift, ensuring that only the relative internal-space structure is retained. To validate the lifting, we compare the reconstructed parallel-space projection, after applying an optimal similarity transformation, with the corresponding physical-space DDQC patch.

\begin{figure}
  \centering
  \includegraphics[width=0.8\textwidth]{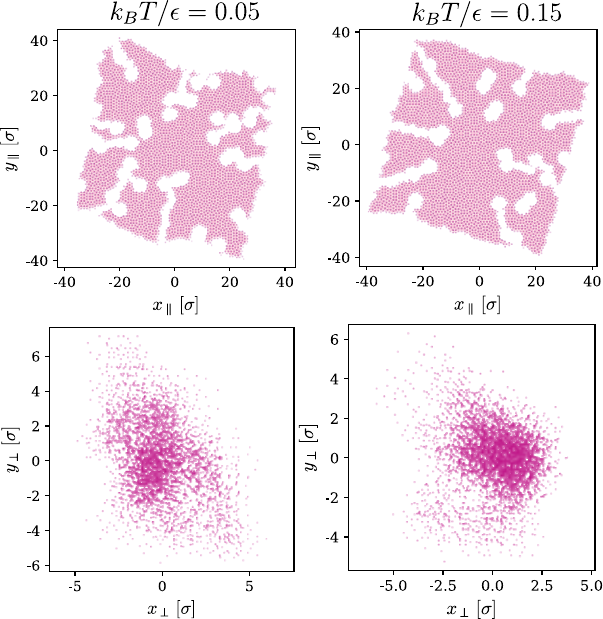}
  \caption{Representative lifted cores used in the static geometry analysis. The parallel-space panels show the scaled projection of the connected, defect-filtered core onto the physical-space basis of the dodecagonal quasicrystal (DDQC), with the reconstructed square-triangle tiling overlaid. The perpendicular-space panels show the corresponding vertices projected into internal space after centering the arbitrary origin of the lift. The compact acceptance-domain structure observed in perpendicular space, together with the well-ordered square-triangle tilings in parallel space, confirms that the protocol-matched samples retain the structural characteristics of well-ordered DDQC square-triangle quasicrystals.}
  \label{fig:supp-lifting-spaces}
\end{figure}

The perpendicular-space coordinates define an internal-space point cloud for each replica.
From this cloud, we extract two static descriptors: a phason-strain length scale $\xi$, obtained from the asymptotic scaling of perpendicular-space distances $d_{\perp}$ with physical-space distances $d_{\parallel}$, and a local outlier score that quantifies how isolated a site is within the perpendicular-space distribution.
Let $\hat{\rho}_{\perp}^{(r)}(\mathbf{x}_{\perp,i})$ denote the kernel-density estimate evaluated at the centered perpendicular-space coordinate of particle $i$ in replica $r$. We define the outlier score as
\begin{equation}
  s_i^{(r)}
  =
  -\ln\!\left[
  \frac{\hat{\rho}_{\perp}^{(r)}(\mathbf{x}_{\perp,i})}
  {\max_{j\in\mathcal{V}_r}\hat{\rho}_{\perp}^{(r)}(\mathbf{x}_{\perp,j})}
  \right],
  \qquad i\in\mathcal{V}_r .
\end{equation}
By construction, $s_i=0$ corresponds to a site located in the densest region of the perpendicular-space cloud, whereas larger values of $s_i$ indicate increasingly isolated sites.  If reversible switching is encoded by static geometric irregularities, one would expect particles with large $s_i$
 to be preferentially associated with subsequent rearrangements.

\begin{figure}
  \centering
  \includegraphics[width=0.7\textwidth]{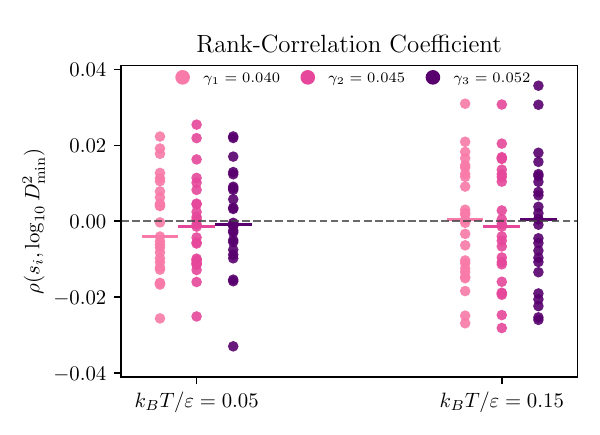}
  \caption{Replica-level rank-correlation coefficient between static perpendicular-space outlier score and branch-resolved rearrangement magnitude. For each protocol-matched replica, we compute the Pearson  coefficient $\rho$ between the static perpendicular-space outlier score $s_i$ and the rearrangement magnitude  $y_i=\log_{10}D^2_{\min}$, evaluated at $\gamma_1$, $\gamma_2$, and $\gamma_3$,  over the defect-filtered lifted core. Each point corresponds to a single replica, while the short horizontal bars denote the median correlation within each temperature-amplitude group. Across all amplitudes and both temperatures, the correlations remain narrowly distributed around zero, indicating that the static perpendicular-space outlier score is not a meaningful   predictor of where reversible rearrangements occur.}
  \label{fig:si-static-rankcorr}
\end{figure}

To compare this static score with the AQS response, we use the continuous affine-corrected peak rearrangement magnitude on each branch $a\in\{\gamma_1,\gamma_2,\gamma_3\}$ of the hierarchical protocol,
\begin{equation}
  y_i^{(r,a)}=\log_{10}D^2_{\min,i}(\gamma_a),
  \qquad i\in\mathcal{V}_r \, ,
\end{equation}
without imposing an activity threshold. The per-replica correlation coefficient shown in Fig.~\ref{fig:si-static-rankcorr} is defined as
\begin{equation}
  \rho^{(r,a)}
  =
  \mathrm{corr}\!\left[
  \mathrm{rank}\!\left(s_i^{(r)}\right),
  \mathrm{rank}\!\left(y_i^{(r,a)}\right)
  \right]_{i\in\mathcal{V}_r},
\end{equation}
i.e. the Pearson correlation computed on  the rank-transformed variables.
This metric directly  probes monotonic associations between static geometry and dynamical response, without binning particles or assuming a linear relationship between the underlying observables.

Fig.~\ref{fig:si-static-rankcorr} shows the central result of this analysis.
For the outer branch $\gamma_3$, the median coefficient is $-9.3\times10^{-4}$ at $k_BT/\varepsilon=0.05$ and $5.4\times10^{-4}$ at $k_BT/\varepsilon=0.15$, with interquartile ranges of $[-5.5,\,8.7]\times10^{-3}$ and $[-10.8,\,10.4]\times10^{-3}$, respectively.
Similar near-zero, sign-indefinite values are obtained for $\gamma_2$ and $\gamma_1$, showing that the absence of correlation is not specific to a particular branch amplitude.
The phason-strain estimates exhibit comparable agreement between temperatures, with $\xi_{\mathrm{rms}}=0.075\pm0.023$ at $k_BT/\varepsilon=0.05$ and $\xi_{\mathrm{rms}}=0.066\pm0.018$ at $k_BT/\varepsilon=0.15$, and substantial overlap between the corresponding distributions.

The protocol-matched dataset therefore reinforces the conclusion that both ensembles are consistent with the same broad random-tiling geometry, while static parallel- and perpendicular-space descriptors fail to identify the particles that act as reversible switching sites.
The memory behavior is thus inherently dynamical and interaction-dependent, rather than being encoded in static perpendicular-space positions.

\section{Eshelby-Like Stress Field}\label{ssec:eshelby}

\begin{figure}
    \centering
    \includegraphics[width=0.7\textwidth]{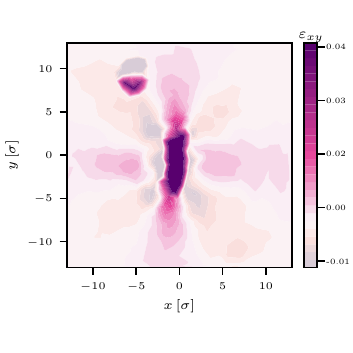}
    \caption{Particle-resolved fitted local shear field around the representative forward tile-switch hysteron discussed in the main text.
    Colors show the local shear component $\varepsilon_{xy}$ in the same core-centered coordinate frame and radial window as the displacement-field panel in Fig.~3c of the main text.
    The color scale is clipped at the $95$th percentile to preserve visibility of  weaker, spatially extended structure in the presence of a localized high-amplitude feature.}
    \label{fig:si-hysteron-stress-panel}
\end{figure}

To construct Fig.~\ref{fig:si-hysteron-stress-panel}, we start from the two energy-minimized configurations immediately before and after the representative A$\to$B switching event. For each particle $i$, we use the neighbor set $\mathcal{N}(i)$ defined by the pre-switch Delaunay graph to determine the best-fit local affine transformation $\mathbf{F}_i$, obtained by minimizing
\begin{equation}
\sum_{j\in\mathcal{N}(i)}\left|\Delta\mathbf{r}_{ij}^{\,\mathrm{post}}-\mathbf{F}_i\Delta\mathbf{r}_{ij}^{\,\mathrm{pre}}\right|^2 \, ,
\label{eq:f_min}
\end{equation}
where $\Delta\mathbf{r}_{ij}=\mathbf{r}_j-\mathbf{r}_i$ is evaluated using the periodic minimum-image convention.
This construction is identical to the local least-squares procedure used to define the non-affine displacement $D^2_{\min}$ in the \emph{Materials and Methods} of the main text, but here it is expressed explicitly in terms of a local deformation map rather than solely through the non-affine residual.
From the fitted tensor, we define the symmetric local strain tensor $\boldsymbol{\varepsilon}_i=(\mathbf{F}_i+\mathbf{F}_i^{\mathsf{T}})/2-\mathbf{I}$ and extract its off-diagonal component $\varepsilon_{xy,i}=(F_{xy,i}+F_{yx,i})/2$.
The resulting field shows that the strongest positive shear is concentrated in a narrow vertical band passing through the switching core, while weaker positive and negative lobes extend into the surrounding quasicrystal.
This provides the particle-resolved shear field underlying the statement in the main text that the event is accompanied by a sign-changing shear redistribution. We emphasize that this panel is intended as a direct visualization of the fitted local shear field for a representative event, rather than as evidence for a universal asymptotic far-field form.

To test the Eshelby-compatible form quoted in the main text, we restrict the analysis to  particles in an annulus surrounding the switching patch, spanning  $r_{\mathrm{in}}\approx 6\sigma$ to $r_{\mathrm{out}}=12\sigma$, thereby focusing on  the surrounding elastic response rather than the strongly non-affine core.
For each particle in this annulus, we compute polar coordinates $(r_i,\theta_i)$ relative to the weighted center of the switching core and evaluate the scaled quantity $r_i^2\varepsilon_{xy,i}$.
If the coarse-grained local shear field follows the idealized far-field form $\varepsilon_{xy}(r,\theta)\propto \cos[4(\theta-\theta_0)]/r^2$, then multiplication by $r^2$ removes the leading radial decay, leaving  a predominantly angular dependence.
We therefore partition the full angular range into $24$ equal bins, discard bins containing fewer than $8$ particles, and compute the average $r^2\varepsilon_{xy}$ within each retained bin. The resulting angular profile is then fitted using weighted least squares to the fourth-order harmonic form $c_0+c_4\cos(4\theta)+s_4\sin(4\theta)$, with each bin weighted by its number of particles.
In this representation, the fitted amplitude quantifies the strength of the four-lobed shear redistribution after removing the expected $1/r^2$ decay, while the fitted phase determines the orientation of the quadrupolar  pattern.
For the representative event shown here, the fit yields $R^2=0.724$, indicating that this simple angular form captures a substantial fraction of the variation in the annulus-averaged $r^2\varepsilon_{xy}$ field within the selected fitting window.
The fit should therefore be interpreted as a compact empirical consistency check for an Eshelby-like quadrupolar shear field surrounding  the switching event, rather than as evidence that the particle-resolved field exactly follows  the corresponding continuum expression.

\section{Absence of Loop Return Point Memory in a Periodic Approximant}\label{ssec:approximant}

The loop return point memory (loop-RPM) reported in the main text is carried by localized, bistable tile-switch
rearrangements. To assess whether these rearrangements rely on  the structural disorder of a
self-assembled quasicrystal or only on its dodecagonal symmetry, we repeat the analysis
on a defect-free periodic crystalline approximant of the dodecagonal tiling. The approximant is constructed as a
random inflationary Stampfli square--triangle tiling of type $[d,I]=[5,2]$, where $d=5$
denotes the size of the initial square lattice and $I=2$ the number of inflation steps applied~\cite{ulugolDefectsEnhanceStability2024}. When embedded in a periodic simulation box, the resulting structure contains
$N=5225$ particles and preserves the local dodecagonal motifs of the quasicrystal, while
remaining spatially periodic and therefore phason-locked. We subject the approximant to the same
interaction potential, density, and athermal quasistatic shear (AQS) protocol used for the quasicrystal, and analyze five
independent realizations.

\begin{figure}
    \centering
    \includegraphics[width=\textwidth]{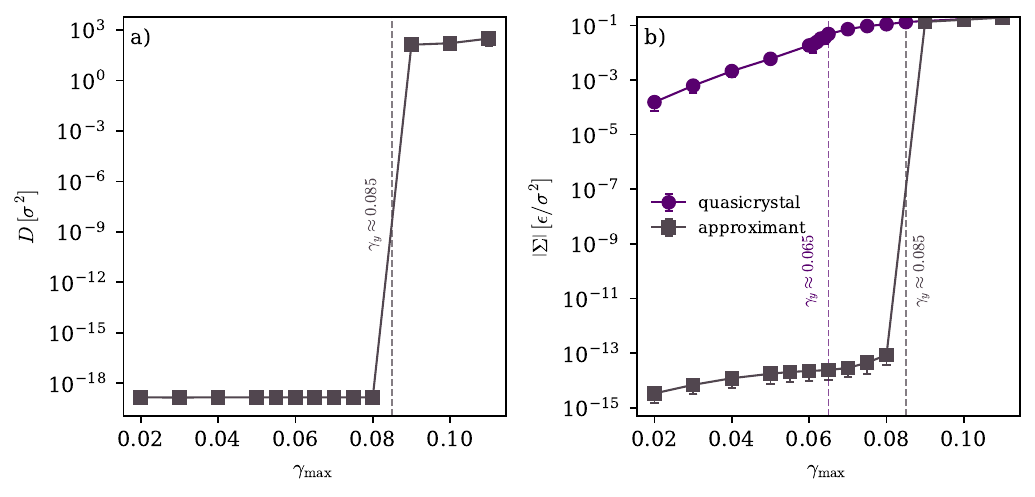}
    \caption{Mechanical response of a periodic crystalline approximant of the DDQC (random
    inflationary Stampfli tiling of type $[d,I]=[5,2]$, $N=5225$, five realizations).
    a)~Stroboscopic diffusion coefficient $D$ as a function of strain amplitude $\gamma_{\max}$. The diffusion coefficient $D$
    remains at the numerical floor below the yielding threshold $\gamma_y\approx0.085$ (dashed
    line) and increases sharply above it. b)~Magnitude of the hysteresis-loop area
    $\left| \Sigma \right| \equiv \oint\sigma_{xy}\,\mathrm{d}\gamma$ per cycle as a function of $\gamma_{\max}$ for both the
    approximant and the self-assembled quasicrystal, computed using identical protocols. Below yielding, the
    quasicrystal develops finite reversible-plastic hysteresis that increases smoothly with
    amplitude, whereas the approximant remains at the numerical floor and exhibits purely elastic response until
     yielding. Dashed lines mark the respective yielding thresholds ($\gamma_y\approx0.065$
    for the quasicrystal and $0.085$ for the approximant). Markers denote averages over the five realizations, with  error
    bars showing the standard deviation, which is smaller than the markers where not visible.}
    \label{fig:approx-no-rpm}
\end{figure}

The approximant exhibits a well-defined yielding transition. Its stroboscopic diffusion
coefficient $D$ remains at the numerical floor for $\gamma_{\max}\lesssim0.085$
and then increases by several orders of magnitude at $\gamma_y\approx0.085$
[Fig.~\ref{fig:approx-no-rpm}a], confirming that it behaves as a genuine yielding solid. We then apply the
nested protocol $\gamma_3\to\gamma_2\to\gamma_1\to\gamma_2\to\gamma_3$ with
$(\gamma_3,\gamma_2,\gamma_1)=(0.07,0.06,0.05)$, chosen relative to the  yielding threshold such that  $\gamma_3/\gamma_y\approx0.82$, closely matching the corresponding quasicrystal value
$\gamma_3/\gamma_y\approx0.80$. Although the pre- and post-excursion stress-strain loops overlap  within
 numerical precision,  this closure reflects simple elastic reversibility rather
than memory formation. The magnitude of the hysteresis-loop area $\left| \Sigma \right| \equiv \oint\sigma_{xy}\,\mathrm{d}\gamma$
remains at the numerical floor  ($\sim\!10^{-14}$) throughout  the entire sub-yield regime and increases abruptly only upon yielding,  where it transitions directly to large irreversible loops  [Fig.~\ref{fig:approx-no-rpm}b].

In contrast, over the same sub-yield  strain amplitudes, the self-assembled quasicrystal develops finite hysteresis arising from  reversible plastic rearrangements. The hysteresis increases smoothly with strain amplitude and is up to $\sim\!10^{11}$ times larger than in  the approximant  
 at the same amplitudes. This finite hysteresis reflects the reversible plastic rearrangements that encode the mechanical memory response. 
 The periodic approximant therefore lacks
the reversible-plastic regime that supports loop-RPM in the quasicrystal. It deforms purely
elastically up to yielding, with no intermediate window of reversible plastic
rearrangements. We attribute this difference to the absence of bistable rearranging units in the phason-locked
approximant. Unlike the quasicrystal, it  lacks the  soft, switchable tile environments created by aperiodic structural disorder, consistent with the established role of  disorder in enabling mechanical memory under cyclic shear~\cite{sethna1993hysteresis,keimMemoryFormationMatter2019}.

\bibliography{bibliography}